\documentclass[review]{elsarticle}

\usepackage{lineno,hyperref}

\journal{Journal of \LaTeX\ Templates}

\begin{document}

\begin{frontmatter}

\title{An Efficient and Wear-Leveling-Aware Frequent-Pattern Mining on Non-Volatile Memory}

\author{Jiaqi Dong, Runyu Zhang, Chaoshu Yang}
\address{College of Computer Science, Chongqing University}


\cortext[mycorrespondingauthor]{Corresponding author}
\ead{jq0783@cqu.edu.cn}


\begin{abstract}
Frequent-pattern mining is a common approach to reveal the valuable hidden trends behind data. However, existing frequent-pattern mining algorithms are designed for DRAM, instead of persistent memories (PMs), which can lead to severe performance and energy overhead due to the utterly different characteristics between DRAM and PMs when they are running on PMs. In this paper, we propose an efficient and Wear-leveling-aware Frequent-Pattern Mining scheme, WFPM, to solve this problem. The proposed WFPM is evaluated by a series of experiments based on realistic datasets from diversified application scenarios, where WFPM achieves 32.0\% performance improvement and prolongs the NVM lifetime of header table by 7.4$\times$ over the EvFP-Tree.
\end{abstract}

\begin{keyword}
\texttt{Frequent-pattern mining}\sep \texttt{NVM}\sep \texttt{Wear-leveling}
\end{keyword}

\end{frontmatter}


\section{Introduction}

Data mining is a highlighted technology to reveal the valuable trends behind the large datasets~\cite{evfptree16,inmemory-bigdata15}. The frequent-pattern mining is an active area of data mining, which is used to identify the frequent-occurring itemsets and patterns in a given dataset~\cite{evfptree16,fpgrowth00}. Several prefix-tree based approaches, such as AFPIM~\cite{Koh2004An}, CFP-tree~\cite{Sucahyo2004CT}, CATS tree~\cite{CATS03}, CanTree~\cite{Leung2007CanTree}, and CP-tree~\cite{Tanbeer2008CP}, have been proposed to optimize the mining tree structure for both space efficiency and performance. However, these frequent-pattern mining algorithms are designed for volatile and energy-inefficient DRAM.

The emerging Non-Volatile Memories (NVMs), such as Phase-Change Memory (PCM)~\cite{palangappa2015wom} and 3D-XPoint~\cite{3DX}, are considered as the promising DRAM replacement for computer main memory. However, they have intrinsic drawbacks in write activities~\cite{chang2015marching}. To our best knowledge, there are only a few NVM-oriented frequent-pattern mining methods, such as EvFP-tree~\cite{evfptree16,liu2017durable} and the parallel version of EvFP-tree (PevFP-tree)~\cite{lin2017scalable}, designed for NVMs. In detail, EvFP-tree proposed the MBA encoding scheme and Lazy Counter to reduce the bits change of Header table while reducing the writes of support counter of the FP-tree nodes. Moreover, PevFP-tree adopts the hash-walk algorithm to reduce the redundant reads on NVM. Accordingly, both EvFP-tree and PevFP-tree can achieve high mining performance by reducing the writes and reads on NVM. However, the MBA encoding scheme still makes low bits wear larger than high bits for an item in the header table. Moreover, they also do not consider the “double writes problem” when the FP-tree construction algorithm sorts the items of transactions. Finally, both EvFP-tree and PevFP-tree failed to fully take advantages of the counted frequencies of items.

In this paper, we present an efficient and Wear-leveling-aware Frequent-Pattern Mining scheme, called WFPM, to solve these problems. The WFPM consists of Wear-leveling-aware Sliding Counter for header table (WSC), Copy-free Growth Mechanism (CGM), and Sorted Hash Walk (SHW), which is used to reduce the write operations on Header Table, reduce the write and read times in FP-tree construction process, respectively. The proposed WFPM is evaluated by a series of experiments based on realistic datasets, where WFPM achieves 32.0\% performance improvement and 30.6\% write reductions compared with EvFP-tree, the state-of-the-art NVM-oriented FP-tree.
The main contributions of this paper are summarised as follows:
\begin{itemize}
  \item The proposed SCW of WFPM adopts a sliding counter scheme to achieve wear-leveling for the header table area of NVM.
  \item The CGM of WFPM is proposed to bypass the sorting process to solve the “double writes problem” of FP-tree construction, which can significantly reduce the write traffics to NVM.
  \item The proposed SHW of WFPM constructs low-overhead ordered linked lists of the hash table to accelerate the FP-tree traversal, which can significantly reduce the read traffics to NVM.
\end{itemize}

The rest of this paper is organized as follows. In Section 2, we introduce the background and discuss the motivation. We present the design and implementation of WFPM in Section 3. The experimental results are detailed in Section 4. Finally, we draw a conclusion in Section 5.

\section{Background and Motivation} \label{Sec:Motivation}

\subsection{Background}
Frequent-pattern mining is an important area of data mining, which is used to discover the valuable trends behind large datasets. To improve the space efficiency of Apriori~\cite{apriori94}, the frequent-pattern tree (FP-tree)~\cite{fpgrowth00} adopts a prefix tree structure to compactly maintain the candidates of frequent patterns. As shown in Figure 1. In the example dataset, each row represents a transaction with its transaction ID (TID) and the associated data items (called itemset).  FP-construct only needs to scan the database twice~\cite{fpgrowth00}. Firstly, FP-construct scans the database to discover all frequent 1-itemsets and sorts these 1-itemsets in order of descending frequency of occurrence in the Header table. Secondly, as shown in Figure 1(c), FP-construct reads transactions in the database and sorts them in descending order.

For the FP-tree construction, in the beginning, only the root node of the FP-tree exists, and the corresponding pattern of the root node is simply an empty pattern, <null>. As the first pattern <a, c, d> is discovered, the corresponding node for the pattern a will be created since no node other than the root node exists in the FP-tree. The support counters of the newly created nodes are initialized to 1, as the corresponding pattern <a, c, d> has occurred for the first time (Figure 1(d)). Likewise, the discovery of the pattern <a, c> creates a new node for the patterns <c>, as shown in Figure 1(d), the support count of node a is 2 and node c is initialized to 1, and so on. We can observe that FP-construct needs to add supporter count in a node when a pattern is being inserted into FP-tree. To overcome these problems, EvFP-tree proposed an MBA encoding scheme to reduce the wear of the Header table on NVM, as shown in Table 1. Moreover, EvFP-tree adopts the lazy counter scheme to reduce the writes on NVM, which removes all unnecessary node updates by postponing the summarizing of support counters until the whole dataset is scanned. Once all transactions have been processed, a depth-first search is carried out to walk the FP-tree to update the support counter of every node as the sum of the present value of the support counter and the support counters of all children of the present node. To improve both the space efficiency and performance of FP-tree traversal, as shown in Figure 2, PevFP-tree proposes a hash-walk algorithm which employs a hash table in each FP-tree node to preliminarily classify the child nodes, so that we do not need to perform an inefficient linear search in all child nodes.

\subsection{Motivation}
Firstly, as mentioned above, we can observe that the classical FP-construct can cause large writes of the Header table on NVM. Meanwhile, the MBA encoding scheme of EvFP-tree still makes low bits wear larger than high bits for an item in the header table. Secondly, the “double writes problem” of sorting items of discovering a transaction cause largely unnecessary writes, which can seriously reduce the performance of frequent-pattern mining. Finally, PevFP-tree uses the hash table to link all child nodes and the sequence of nodes according to the emergency order instead of occurrence. Accordingly, the hash-walk algorithm still leads to large unnecessary read times on NVM.


\section{Design and implementation}\label{Sec:DesignFramework}

\subsection{The Wear-leveling-aware Sliding Counter}
We propose wear-leveling-aware sliding counters to substitute for the vulnerable regular counters in the header table. As illustrated in Figure 3, we divide the 64-bit area into a 4-bit metadata region and a 60-bit sliding region. Figure 3 (a) shows the initial state of a sliding counter, of which the highest counting block is the 8th block in the counting region. When  , each counting block will be moved to their left block. The highest counting block will finally reach the border of the counting region, as shown in Figure 3 (b). As shown in Figure 3 (c), we invert the direction of these counting blocks. To be specific, the serial number of the lowest counting block is 7, while that of the highest counting is 14. For i from 1 to 4, we swap each pair of (i+6)th and (15-i)th counting blocks. The direction bit is set to 0, indicating the direction of movements is to the right. As a consequence, the packing order of these counting blocks has reversed, in which the high blocks occupy the previous low blocks and vice versa. Subsequently, as shown in Figure 3 (d), the counting blocks move to the right as the counter value increases, opposite to the previous direction of movement.


\subsection{The Copy-free Growth Mechanism}
The WFPM bypasses the sorting process to conduct insertions without extra write activities. As illustrated in Figure 4, WFPM traverses the header table in descending order to fetch the order of items in transactions. When encountering an item that also appears in the transactions, we search down the tree from root to keep track of the pattern. Once inserting a new item into the tree, we create a new node with an initialized sliding counter. After processing the whole transaction, we only add one to the value of the last sliding counter, akin to the lazy counter mechanism in EvFP-Tree.


\subsection{The Sorted Hash Walk}
The PevFP-Tree has failed to fully take advantages of the counted frequencies of items. We construct a sorted hash walk mechanism to minimize the overhead of traversing the pointers in hash walks. As demonstrated in Figure 5, when inserting a new item pointer to the linked list in a hash table, we compare the frequency of the inserted item with that of items in the linked list. Finally, the new item pointer will be linked just after the older one with a larger frequency. Consequently, the item pointers are linked in descending order according to their frequencies, which provides fast accesses to most frequent items in the search down process.

\section{Evaluation}\label{Sec:Exp}


\subsection{Experimental setup}
We have implemented traditional EvFP-Tree and WFPM for experiments. We have also integrated the hash walk mechanism into EvFP-Tree for fairness. To evaluate the efficacy of the WFPM, we perform a series of trace-driven simulations based on publicly-available practical datasets from several repositories~\cite{fimi15,icsuci-archive15,dense-sparse-datasets}. The experiments are conducted on a workstation running RedHat Linux 6.0 with two eight-core processors clocked at 2.6 GHz and 64 GB of DRAM. We assumed that the PCM is used in the experiments to store the FP-trees. The performance statistics of the PCM are obtained from~\cite{hu2013software}, where a PCM read and write (SET/RESET) take 6.82 ns and 152.20/12.20 ns of time, respectively. As for energy consumption, a PCM read and write (SET/RESET) operation takes 64 pJ and 70.0/876 pJ, respectively. To simulate the resource-limited embedded environment, we also assumed a 32-KB, four-way associative cache with the least-recently used (LRU) replacement policy~\cite{lrfu01} is available. An SRAM read or write operation takes 1.41 ns~\cite{hu2013software}.

\subsection{Experimental Results and Discussion}
This subsection will conduct comparisons in terms of NVM write activities from various aspects. We first compare the times of NVM write activities to demonstrate the mitigation of imbalanced wears on counters in the header table. Figure 6 shows the maximum bit flips of counters in the header table. In this figure, the “Total” columns represent the numbers of write activities on counters, which equals to the times of last-bit flips in the traditional FP-Tree. We can observe from Figure 6 that the WFPM gains 1.9$\times$~12.5$\times$ reductions on the maximum bit flips compared with EvFP-Tree. Moreover, the times of reductions increase with the total number of write activities. These results highlight the effectiveness of the proposed wear-leveling-aware sliding counters.

The total times of NVM write activities are shown in Figure 7. Benefit from the copy-free growth mechanism, our WFPM achieves 10.4\%~41.9\% reductions on total write activities compared with EvFP-Tree. This is because the WFPM avoids the sorting process of transactions, and thereby eliminates the double write of each transaction. Figure 8 shows the total times of NVM read activities of two schemes. WFPM reduces 7.6\%~85.3\% NVM reads compared to EvFP-Tree. This improvement is mainly attributed to the sorted hash walk mechanism. The WFPM intrinsically decreases the times of read activities in the search down process, which form the majority of NVM reads.

Figure 9 and Figure 10 show the total elapsed time and energy consumption of WFPM and EvFP-Tree. We can observe that WFPM constantly outperforms EvFP-Tree with all realistic workloads in terms of timing performance and energy consumption. Specifically, WFPM achieves 12.8\%~41.3\% reductions on total elapsed time and 14.3\%~40.9\% reductions on energy consumption compared to EvFP-Tree, respectively. These results demonstrate the efficiency of WFPM.



\section{Conclusion}\label{Sec:Con}
We have conducted in-depth investigations on existing FP-Trees and revealed the drawbacks of them deployed on NVMs. We have proposed efficient mechanisms to address these issues. Comprehensive evaluations have shown that our scheme significantly outperforms EvFP-Tree from various aspects.

%

\section*{References}

\bibliography{ref}

\end{document}